\documentclass[12pt,epsf,epsfig,psfig]{article}
\usepackage{graphicx}
\usepackage{epsfig}
\usepackage{cite}
\def\e{\epsilon}

\def\be{\begin{equation}}
\def\ee{\end{equation}}

\def\lsim{\raise0.3ex\hbox{$<$\kern-0.75em\raise-1.1ex\hbox{$\sim$}}}
\def\gsim{\raise0.3ex\hbox{$>$\kern-0.75em\raise-1.1ex\hbox{$\sim$}}}

\def\NP{{ Nucl.\ Phys.\ }}
\def\PL{{ Phys.\ Lett.\ }}
\def\PR{{ Phys.\ Rev.\ }}

\def\PRL{{ Phys.\ Rev.\ Lett.\ }}

\def\EP{{ Europ.\ Phys.\ J.\ C}}

\begin{document}

~

\bigskip

\centerline{\Large \bf High Energy Hadron Production}

\bigskip

\centerline{\Large \bf as Self-Organized Criticality}

\vskip1cm

\centerline{\large \bf Paolo Castorina$^{a,b}$ and Helmut Satz$^c$}

\bigskip

\centerline{a: INFN sezione di Catania, Catania, Italy}

\centerline{b: Faculty of Physics and Mathematics, Charles University, Prague,
Czech Republic}

\centerline{c: Fakult\"at f\"ur Physik, Universit\"at Bielefeld, Germany}

\vskip1.6cm

\centerline{\large \bf Abstract}

\medskip

In high energy nuclear collisions, production rates of light nuclei as 
well as those of hadrons and hadronic resonances agree with the 
predictions of an ideal gas at a temperature $T=155 \pm 10$ MeV. In an 
equilibrium hadronic medium of this temperature, light nuclei cannot survive. 
We propose that the observed behavior is due to an evolution in global 
non-equilibrium, leading to self-organized criticality. At the confinement 
point, the initial quark-gluon medium becomes quenched by the vacuum, breaking 
up into all allowed free hadronic and nuclear mass states, without formation 
of any subsequent thermal hadronic medium.

\vskip1.3cm

The production rates of hadrons and hadronic resonances in high energy 
collisions, from $e^+e^-$ to $A-A$, have for some years been somewhat
enigmatic. The relative yields are found to agree with those obtained
from an ideal resonance gas \cite{resgas} at the (pseudo)critical confinement 
temperature $T_c = 155 \pm 10$ MeV \cite{baza1,baza2}; with an additional 
parameter speci-fying an overall production volume $V$ \cite{alice}, so are 
the absolute yields\footnote{At lower energies, additional parameters for
baryon density and strangeness suppression may be needed.}. The conventional
interpretation assumes that the collision leads to a hot quark-gluon plasma,
which evolves, cools and at $T_c$ undergoes a transition to a hadron gas of 
that temperature. This interacting hadron gas then expands, cools further, 
and eventually freezes out into free hadrons.
The curious feature is that the hadron abundances are already specified 
once and for all at $T_c$ and are not subsequently modified in the evolution
of the hadron gas. In general, the interactions are taken to cease completely 
only at a kinematic freeze-out at a somewhat lower temperature.
The ideal gas description at $T_c$ is found to be valid even for resonances 
such as the $\rho$ or the $\Delta$, which are in principle easy to break up
in a hadron gas of 155 MeV temperature. To maintain for hadron production
in nuclear collisions both the ideal hadron
gas abundances as given at $T_c$ and the existence of a thermal hadronic 
medium below the critical point, one has taken recourse to various features: 
a very rapid decrease of the hadron gas density, very weak interactions 
through hadronic collisions, a very short life-time of the hadron gas, 
and more (see e.g. \cite{ex1}). In elementary collisions 
($e^+e^-$ or $p-p$), hadron production has been accounted for by quantum 
mechanisms leading to behavior of a thermal form, without implying a
subsequent thermal hadronic medium \cite{ex2,ex3,ex4,
ex5}

\medskip

The mentioned enigma was further enhanced by recent LHC data of $Pb-Pb$
collisions at $\sqrt s = 2.76$ TeV, taken by
the ALICE collaboration \cite{alice}. It is observed that even the yields
for light nuclei, deuteron, 3Helium, hyper-triton, 4Helium and
their antiparticles, are in accord with a formation temperature of 155 MeV.
Since these states have binding energies of a few MeV and are generally
of larger than hadronic size, their survival in the assumed hot hadron gas
poses an even more striking puzzle, often characterized as ``snowball in 
hell'' \cite{stachel}. In a situation of approximate global equilibrium, with
much slower evolution, as presumably existed in the nucleosynthesis after 
the Big Bang, the rate of deuteron and helium production has in fact nothing 
to do with the hadronization temperature: it is determined by the binding 
energies of the nuclei. Even if in the early universe the great abundance of 
photons can enhance the break-up of possible nuclei, an interacting hadron gas 
with many pions would also do this. Our basic question thus is:
why are the yields for the production of light nuclei determined by the 
rates as specified at the critical hadronization temperature, although in hot 
hadron gas they would immediately be destroyed? 

\medskip

In the present note, we want to show that a solution to this puzzle can be
obtained by abandoning the idea of a thermal hadron medium existing below
the confinement point. Instead, we propose that the hot quark-gluon system, 
when it cools down to the hadronization temperature, is effectively quenched 
by the cold physical vacuum. The relevant basic mechanism for this is 
self-organized criticality, leading to universal scale-free behavior, as 
in fact also obtained in many other cases. We first recall the corresponding 
scenario.

\medskip

Self-organized criticality (SOC) \cite{Bak} 
is the evolution of a non-equilibrium system to
a critical attractor, driven by the individual interactions, without any
tuning of external parameters (for surveys, see \cite{Jensen,Sornette,GP}).
At the critical point, the system becomes
scale-free, so that components of all sizes are accounted for by the same law.
While equilibrium systems require the tuning of thermal parameters 
(temperature, density) to reach critical behavior, non-equlibrium systems
subject to SOC reach the critical point through interactions within the
system itself.

\medskip

The appearance of scale-free behavior at criticality is readily seen by 
considering the correlation function $\Gamma(r)$ of a many-body system, 
generally taken to have the form
\be
\Gamma(r,T) \sim {\exp{-[r/\xi(T)]}\over r^p},
\label{1}
\ee
where $r$ denotes the separation distance of two constitutents and $\xi(T)$
the correlation length in the system at temperature $T$.
The exponent $p$ is equal to unity in the conventional Ornstein-Zernike
formulation in three space dimensions; the general form in $d\geq 3$
space dimension and with the fractal extension $\eta$ \cite{Fisher-eta}
is $p=d-2+\eta$. In most cases, $\eta$ is small or vanishes.
The correlation length $\xi(T)$
defines a temperature-dependent scale specifying at what separation
constituents still are in touch. At the critical point $T=T_c$ of a
continuous phase transition, $\xi \to \infty$, so that there is no longer
a scale-dependent parameter measuring the role of different separations.
The correlation function now shows universal power-law behavior, 
\be
\Gamma(r,T_c) \sim r^{-p}
\label{1n}
\ee
for all separations $r$, and self-organized criticality means 
that the system is governed by such a scale-free form. As a corollary, 
the corresponding susceptibility in the three-dimensional case
\be
\chi(T) = {1\over kT} \int d^3 r \Gamma(r,T) \to  
{4\pi\over kT}\int dr r^{1-\eta}
\label{1a}
\ee
will diverge at the critical point.

\medskip

\begin{figure}[htb]
\centerline{\psfig{file=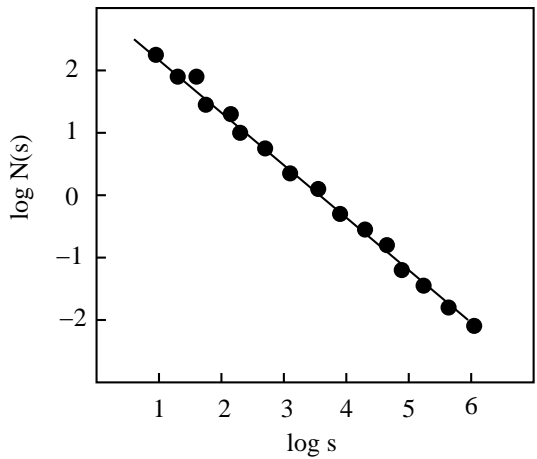,width=7cm}} 
\caption{Distributions of the number $N(s)$ of avalanches vs. their size
$s$}
\label{soc1}
\end{figure}

The typical illustration of SOC proposed in the pioneering work \cite{Bak} 
is the behavior of sandpiles. Pouring sand onto a flat surface leads to a 
pile increasing in size up to a point where the pile has a certain 
critical slope. The addition of further
sand now results in avalanches of various sizes, preventing an overall
increase of the slope. The number $N(s)$ of 
different avalanches of size $s$ 
observed over a long period is found to vary as a power of $s$. Such power-law
behavior 
\be
N(s) = \alpha s^{-p}
\ee
implies 
\be
\log N(s) = A - p \log s, A=\log \alpha
\label{3}
\ee
as shown in Fig.\ \ref{soc1}. Large avalanches are thus governed by the same 
law as small ones - the phenomenon is scale-free. This breaks down only at
the ends, with single grains of sand and the entire pile.

\medskip

Such behavior is in fact found in a variety of situations, from sandpiles to
earthquakes to partitioning of integers. Much of the complexity of our world
may well result from self-organized criticality.
In high energy physics, it is 
immediately reminiscent of the statistical bootstrap model
of Hagedorn \cite{hagedorn}, who had ``fireballs composed of fireballs, which
in turn are composed of fireballs, and so on''. Such scale-free  
composition cascades of massive hadronic states arise also in the 
dual-resonance model \cite{DRM1,DRM2,DRM3}. The general pattern
has been shown to be due to an underlying
structure analogous to the partitioning of an integer into integers 
\cite{blan}, the well-known {\sl partitio numerorum} problem of number theory 
\cite{hardy}, and in fact sand piles and integers were shown to result in 
similar structural patterns \cite{goles}. This in turn has immediate 
consequences on hadron production observed in high energy collisions,
as we shall show. 

\medskip

\begin{figure}[htb]
\centerline{\psfig{file=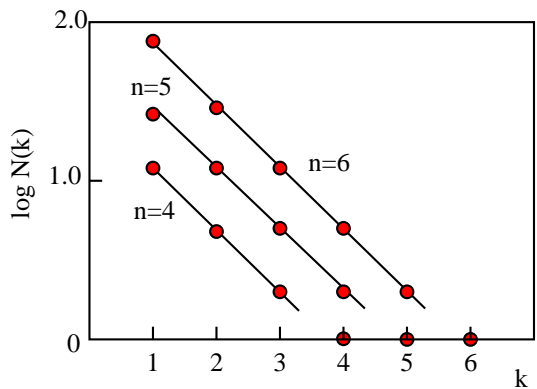,width=7cm}} 
\caption{Distributions of the number $N(k,n)$ of integers $k$ for $n=4,5$ 
and 6}
\label{parti}
\end{figure}

The perhaps simplest form of self-organized criticality is provided by 
partitioning integers. Consider the {\sl ordered} partitioning of an integer 
$n$ into integers. The number $q(n)$ of such partitionings is for $n=3$
equal to four: 3, 2+1, 1+2, 1+1+1, i.e., $p(3)=4$. It is easily shown 
\cite{blan} that in general
\be
q(n) = 2^{n-1} = {1\over 2} \exp\{n \ln 2\}.
\label{p1}
\ee
The problem of unordered partitions is more difficult and only solved 
asymptotically \cite{hardy}. In the ordered case considered above, one thus 
finds that the 
number of partitions increases exponentially with the size of the integer. 
Given an initial integer $n$, we would now like to know the number $N(k,n)$
specifying how often a given integer $k$ occurs in the set of all 
partitionings of $n$. To illustrate, in the above case of $n=3$, we have 
$N(3,3)=1$, $N(3,2)=2$ and $N(3,1)=5$. To apply the formalism of 
self-organized criticality, we have to attribute a strength $s(k)$ to each
integer. It seems natural use the number of partitions for this, i.e., set
\be
s(k) = q(k) = {1\over 2} \exp\{k \ln 2\}.
\label{p2}
\ee
The desired number $N(k,n)$ in a scale-free scenario is then given by
\be
N(k,n) = \alpha(n) [q(k)]^{-p},
\label{p3}
\ee
leading to
\be
\log~\! N(k,n) = -[p \log e\ \ln 2]~\!k + p \log 2 + \log \alpha(n) 
\label{p4}
\ee
as counterpart of eq.\ (\ref{3}). For small values of $n$, $N(k,n)$ is readily
obtained explicitly. In Fig.\ \ref{parti} we thus see that
relation (\ref{p4}) is in fact well satisfied already for $n=4,~5$ and 
$6$, except for slight deviations in the limit $k=n$.
In particular, we find $[p \log e \ln 2] \simeq 0.38$, so that the critical
exponent becomes $p \simeq 1.26$. The appearence of a specific integer $k$
in the set of all partitions of $n$ thus corresponds to the appearence of
an avalanche of a given size in the average over a long time period of the
sandpile case.
 
\medskip

Hagedorn's bootstrap approach \cite{hagedorn} proposes that a hadronic
state of overall mass $m$ can be partitioned into structurally similar
states, and so on. If these states were at rest, the situation would be
identical to the above partioning problem. Since the constituent fireballs 
have an intrinsic motion, the number of states $\rho(m)$
corresponding to a given
mass $m$ is determined by the bootstrap equation
\be
\rho(m) = \delta(m\!-\!m_0) ~+
\sum_N {1\over N!} \left[ {4 \pi \over 3(2\pi m_0)^3} \right]^{N-1} 
\hskip-0.2cm \int \prod_{i=1}^N ~[dm_i~ \rho(m_i)~ d^3p_i] 
~\delta^4(\Sigma_i p_i - p),
\label{bootstrap}
\ee
with $m_0$ denoting the lowest possible mass (the single grain of sand).
The equation can be solved analytically \cite{nahm}, giving
\be
\rho(m) \sim m^{-a} \e^{m/T_H} 
~\to~ \ln \rho \sim {m\over T_H}
- a \ln m, 
\label{exp}
\ee
and $T_H$ as solution of 
\be
\left({2\over 3 \pi}\right)\left(T_H \over m_0\right) 
K_2(m_0/T_H) = 2 \ln 2 - 1,
\label{bs}
\ee
where $K_2(x)$ is a Hankel function of pure imaginary argument. For
$m_0 = m_{\pi} \simeq 130$ Mev, this leads to the Hagedorn temperature
$T_H \simeq 150$ MeV, i.e., to approximately the critical hadronization 
temperature found in statistical QCD. The cited solution gave $a=3$,
but other exponents have also been discussed \cite{F,CP,HS}. 
 
\medskip

The form of eq.\ (\ref{exp}) is an asymptotic solution of the bootstrap
equation; it evidently diverges for $m\to 0$ and must be modified for 
small masses. Using a similar result 
for $\rho(m)$ obtained in the dual resonance model \cite{huang-wein}, 
Hagedorn proposed \cite{yellow} 
\be
\rho(m) = {\rm const.}(1+ (m/\mu_0))^{-a} \exp(m/T_H) 
\label{exp1}
\ee   
where $\mu_0 \simeq 1 - 2$ GeV is a normalization constant.

\medskip

At this point we should emphasize that the forms (\ref{3}), (\ref{p4})
and (\ref{exp}) are entirely due to the self-organized nature of the 
components, with an integer consisting of integers, a fireball of fireballs.
They are in no way a result of thermal behavior. We have expressed
the slope coefficient of $m$ in eq.\ (\ref{exp}) in terms of the Hagedorn 
``temperature'' only in reference to subsequent applications. In itself, it
is totally of combinatorical origin. It is of course possible to
construct a thermodynamics of integer partitions \cite{blan}, with an entropy 
$S(n)=\ln q(n) = n \ln 2 - \ln 2$, leading to a temperature 
$\Theta = dS(n)/dn = \ln 2$. In that sense, the integers then form a
gas of partitions at the critical temperature $\Theta$.

\medskip

We now want to apply the formalism of self-organized criticality to strong
interaction physics. There exists some early work in that direction, in which 
it was argued that Reggeon field theory \cite{gribov} in fact shows such
behavior, with critical exponents in the same universality class as a 
specific avalanche model \cite{FT1,FT2}. In the framework of QCD as basic 
theory of strong interactions, numerical lattice studies have shown that the
deconfinement/confinement transition is in fact a rapid cross-over 
rather than a genuine thermodynamic phase transition \cite{cross}. In the 
chiral limit of two flavor QCD with vanishing quark masses $m_q$, one does 
recover critical behavior \cite{PW,RW}; there is 
a continuous transition at $T=T_c$, with the chiral condensate 
$M=< \psi \bar\psi>$ as order paramenter,
\be
M(T) \sim (T_c-T)^{\beta}, ~ T \leq T_c
\label{ch1}
\ee
in terms of the critical exponent $\beta$. Thus $T_c$ is defined as the 
temperature point at which for $m_q=0$ the chiral condensate $M(T)$ vanishes,
$M(T = T_c,m_q=0)=0$. For a system at $T=T_c$ with $m_q \to 0$, one finds
singular behavior,
\be
M(T_c) \sim m_q^{1/\delta},
\label{ch2}
\ee
with the critical exponent $\delta$. For $m_q=0$, the correlation length
characterizing fluctuations of the chiral condensate diverges as
\be
\xi(T,m_q=0) \sim |T-T_c|^{-\nu},
\label{ch3}
\ee
in terms of the critical exponent $\nu$. 

\medskip

\begin{figure}[htb]
\centerline{\psfig{file=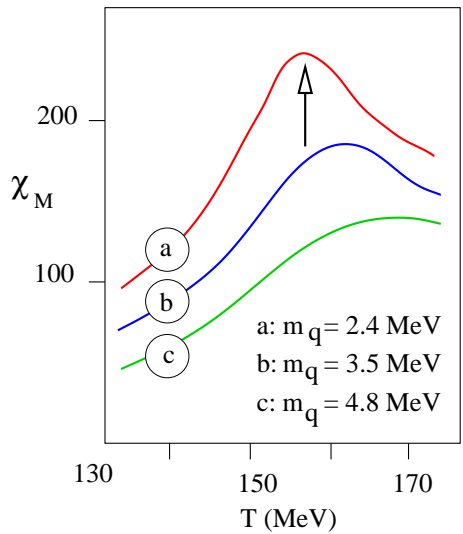,width=6cm}} 
\caption{The temperature dependence of the chiral susceptibility
for three different light quark masses \cite{lahiri}.}
\label{chiral}
\end{figure}

The small but finite $u$ and $d$ quark masses $m_q$ in physical QCD act 
like a weak external field in spin system (see eq.\ (\ref{ch2})), 
preventing genuine singular behavior \cite{laer}. 
The behavior of the system 
for the actual small quark mass values is nevertheless thought to be 
strongly influenced by the near-by singularity. As a result, specific 
thermodynamic variables are sharply peaked, defining a pseudo-critical point 
$T_{pc}(m_q)$ close to $T_c(m_q=0)$. The fluctuation-dissipation theorem
relates the correlation length to the chiral susceptibility  
$\chi_M \sim \partial M(T)/ \partial m_q$, and in Fig.\ \ref{chiral}
this is seen to show a pronounced peak in temperature \cite{lahiri},
which is taken as the pseudocritical temperature of QCD.

\medskip

\medskip

This is now to be applied to high energy nuclear collisions. The picture we
have in mind assumes a sudden quench of the partonic medium produced
in the collision. The initial hot system of deconfined quarks and 
gluons (also interpreted as the molten color glass \cite{larry})
rapidly expands and cools; while this system is presumably in local
thermal equilibrium, the difference between transverse and longitudinal
motion implies a global non-equilibrium behavior. The longitudinal
expansion quickly drives the system to the hadronisation point, and it is 
now suddenly thrown into the cold physical vacuum. The process is not 
unlike that of a molten metal being dumped into cold water. In this quenching
process, the system freezes out into the degrees of freedom presented by 
the system at the transition point and subsequently remains as such, apart
from possible hadron or resonance decays. There
never is an evolving warm metal. In other words, in our case there is no hot 
interacting hadron gas. To obtain that, we would have to adiabatically
lower the temperature of a closed quark-gluon system; it is the sudden 
immersion into the vacuum that causes the quench. The snowball is one of the
allowed states of the system, and so it can appear at the quenching point.
Subsequently, however, it is not in hell, but together with all other
fragments, it finds itself freestreaming in the cold physical vacuum. Whatever
thermal features are observed, such as radial or elliptic hydrodynamic
flow, must then have originated from local equilibrium in the earlier 
deconfined stage \cite{localSOC1,localSOC2}.
The mechanism driving the system rapidly to the critical point
is the global non-equilibrium due to the longitudinal motion provided
by the collision. 

\medskip

In such a scenario, high energy nuclear collisions lead to a system 
which at the critical point breaks up into
components of different masses $m$, subject to self-similar
composition and hence of a strength $\rho(m)$
as given by the above eq.\ (\ref{exp1}). In the self-organized criticality
formalism, this implies that the interaction will produce
\be
N(m) = \alpha [\rho(m)]^{-p}
\label{p5}
\ee
hadrons of mass $m$. With $\rho(m)$ given by eq.\ (\ref{exp1}), the 
resulting powerlaw form
\be
\log N(m) = -m \left({p \log e \over T_H} \right) \left[1 
- \left(a T_H \over
m \right ) 
{\ln(1+{m \over \mu_0})}\right]
+ {\rm const.} 
\label{p6}
\ee
is found to show a behavior similar to that obtained from 
an ideal resonance gas in equilibrium. We emphasize that it 
is here obtained assuming only scale-free
behavior (self-organized criticality) and a mass weight determined by
the number of partitions. No equilibrium thermal system of any kind is assumed.

\medskip

\begin{figure}[htb]
\centerline{\psfig{file=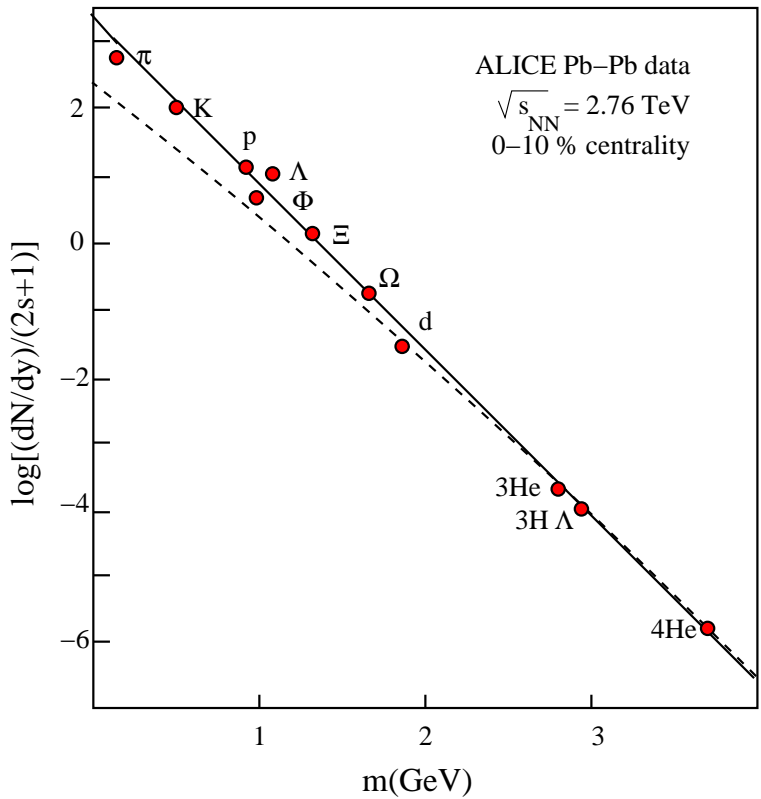,width=10cm}} 
\caption{Yield rates of species at central 
rapidity vs. their mass $m$ \cite{alice}. The solid line corresponds to
eq.\ (\ref{p7}), the dashed line to eq.\ (\ref{p7a}).}
\label{soc3}
\end{figure}

We now consider the mentioned ALICE data \cite{alice}. In Fig.\ \ref{soc3}
the production yields for the different mass states in central
$Pb-Pb$ collisions at $\sqrt s = 2.76$ GeV are shown; in each 
case, the yield is divided by the relevant spin degeneracy. We see that the
yields show essentially powerlike behavior, and the light nuclei follow the 
same law as the elementary hadrons. The solid line 
in Fig.\ \ref{soc3} shows the behavior obtained from eqs.\ 
(\ref{p6}), ignoring for the moment the second term in the square brackets,
\be
\log[(dN/dy)/(2s+1)] \simeq -m\left({0.43~\!p\over T_H}\right) + A,
\label{p7}
 \ee
with $T_H=0.155 MeV$ and fit values $p=0.9$, $A=3.4$. 
The form is evidently in good agreement with the data.

\medskip

Including the correction term to linear behavior that we had omitted above, 
we have
\be
\log[(dN/dy)/(2s+1)] \simeq -m\left({0.43~\!p\over T_H}\right) + 
+ p~\!a\log[1+(m/\mu)] + A,
\label{p7a}
 \ee
The additional term is, as indicated, rather model dependent. It will 
effectively turn the yield curve down for decreasing masses. This is in fact 
necessary, since the decay of heavier resonances will enhance the direct 
low mass meson yields.
To illustrate the effect of the term, we choose $a=3$, corresponding to
the mentioned solution (\ref{exp}) of the bootstrap equation \cite{nahm},
and $\mu=2$ GeV for the normalization. The result is included in Fig.\
\ref{soc3}.

\medskip 

\medskip

Our main result, eq.\ (\ref{p7a}), is quite similar to the result of the
yield calculation in the resonance gas model \cite{resgas}. We have here 
obtained it, however, without the assumption that the confinement transition 
produces a hot interacting resonance gas. Instead, the yield values arise 
in the sudden self-organized critical quench which takes place when the hot 
partonic system  hits the cold vacuum.
We thus conclude that the agreement of high energy hadron production
yields with an ideal resonance gas model does not establish that such 
collisions produce a thermal hadronic medium for $T < T_c$. The observation 
that the yields of light nuclei also agree with the same pattern in fact 
throws serious doubt on the existence of such a medium. In the global
non-equilibrium scenario proposed here, any state, also still heavier nuclei, 
can arise in the quench.

\medskip

As a caveat, we note, however, that the argumentation based on the 
statistical bootstrap model result (\ref{exp1}) is inherently of a qualitative 
nature. The form (\ref{exp1}) itself holds in the limit of large $m$; 
moreover, to obtain it, the discrete hadron spectrum was replaced by a 
continuum. Furthermore, the isospin, strangeness and baryon number
structure of the spectrum are not taken into account. We can therefore
expect at best qualitative agreement with the data - in detail, our
simple model cannot be expected to reproduce the results of a full ideal 
resonance gas analysis and is meant mainly to illustrate the SOC approach.
Further work on a more detailed SOC analysis is in progress.

\medskip

As final comment, we recall the behavior of proton-nucleus or 
nucleus-nucleus collisions at low energy, leading to what is denoted as 
nuclear multifragmentation \cite{multi1,multi2}. The result of such 
collisions will be nuclear 
fragments of size $A$, and the distribution of these fragments is found to 
obey the so-called Fisher law \cite{fisher}
\be 
P(A) = {\rm const.} A^{-\tau},
\label{p8}
\ee     
corresponding to the critical point of droplet condensation, with $\tau =
2.33$. It thus constitutes another instance of self-organized criticality,
with all fragments governed by the same law. The difference between this
form and the one for high energy collisions is that at low energy, the
the break-up is simply into mass fragments, whereas at high energy it produces
all possible excitation states.

\bigskip 

\centerline{\bf Acknowledgements}

\bigskip

It is a pleasure to thank P.\ Braun-Munzinger (GSI),
F.\ Karsch (Bielefeld), D.\ Kharzeev (Stony Brook)
and J.\ Randrup (LBL Berkeley) for stimulating and helpful comments. We are
very grateful to A.\ Andronic (M\"unster/GSI) for help with the data.

\end{document}